\RequirePackage[2020-02-02]{latexrelease}
\documentclass[twocolumn,aps]{revtex4-2}
\usepackage{subfigure}
\usepackage{graphics}

\usepackage{graphicx}
\usepackage[T1]{fontenc}
\usepackage[latin9]{inputenc}
\usepackage{color}
\usepackage{ifsym}
\usepackage{wasysym}
\usepackage{babel}
\usepackage{xr}
\usepackage[normalem]{ulem}
\usepackage{soul}
\newcommand{\beginsupplement}{                     \setcounter{table}{0}
	\renewcommand{\thetable}{S\arabic{table}}                       \setcounter{figure}{0}
	\renewcommand{\thefigure}{S\arabic{figure}}
}

\begin{document}	
	\title{Re-laminarization of elastic turbulence}
	\author{M. Vijay Kumar$^{1,2,3*}$, Atul Varshney$^{1,4,5*}$, Dongyang Li$^{1,6}$, and Victor Steinberg$^{1,2}$}
	\affiliation{$^1$Department of Physics of Complex Systems, Weizmann Institute of Science, Rehovot, Israel 76100\\$^2$The Racah Institute of Physics, Hebrew University of Jerusalem, Jerusalem 91904, Israel\\$^3$Department of Mechanical and Industrial Engineering, University of Illinois at Chicago, Illinois 60607, USA\\$^4$School of Physical Sciences, National Institute of Science Education and Research, HBNI, Jatni-752050, Odisha, India\\$^5$Institute of Science and Technology Austria, Am Campus 1, 3400 Klosterneuburg, Austria\\$^6$College of Nuclear Science and Technology, Harbin Engineering University, Harbin 150001, China}

	\begin{abstract}
            We report frictional drag reduction and a complete flow re-laminarization of elastic turbulence (ET) at vanishing inertia in a viscoelastic channel flow past an obstacle. We show that intensity of observed elastic waves and wall-normal vorticity correlate well with the measured drag above the ET onset.   Moreover, we find that the elastic wave frequency grows with Weissenberg number, and at sufficiently high frequency it causes  decay of the elastic waves, resulting in ET attenuation and drag reduction.  Thus, this allows us to substantiate a physical mechanism, involving interaction of elastic waves with wall-normal vorticity fluctuations, leading to the drag reduction and re-laminarization phenomena at low Reynolds number.
	
\end{abstract}

\maketitle
Curvilinear flows of dilute polymer solution exhibit elastic instabilities and elastic turbulence (ET) at vanishing inertia. In such flows, the elastic stress generated by polymer stretching along curved streamlines initiates a back-reaction on the flow 
in the curvature direction  and triggers a linear normal mode elastic instability and ET at $Wi\gg1$ and $Re\ll1$ \cite{steinberg_elastic_2021, groisman_elastic_2000}. However, this instability mechanism becomes ineffective in zero curvature parallel shear flows, such as pipe, channel and plane Couette \cite{larson_instabilities_1992, shaqfeh_purely_1996, steinberg_elastic_2021, groisman_elastic_2000,pakdel_elastic_1996, varshney_elastic_2017}. Here, the degree of polymer stretching \cite{bird_dynamics_1987} is defined by the Weissenberg number $Wi=\lambda U/d$ and the ratio of inertial and viscous stresses by the Reynolds number $Re=\rho Ud/\eta$; $U$ is the mean fluid speed, $\lambda$ is the longest polymer relaxation time, $\eta$ and $\rho$ are the solution viscosity and density, respectively, and $d$ is the characteristic vessel size.

The first signature of a linear  elastic instability was reported in a viscoelastic channel flow obstructed by two widely spaced obstacles \cite{varshney_elastic_2017}. The linear Hopf instability occurs due to a breaking of time-reversal symmetry  leading to span-wise oscillations of a pair of eddies. Further, two shear (mixing) layers with zero curvature streamlines are generated due to the vortex pair elongation with $Wi$ \cite{varshney_mixing_2018}. It follows by a secondary transition directly to ET and elastic waves at $Re\ll1$ \cite{varshney_mixing_2018, varshney_drag_2018,varshney_elastic_2019}. The measured friction factor, $f/f_{lam}$,  as a function of $Wi$ exhibits distinct scaling exponents in three flow regimes: $Wi^{0.5}$ (transition), $Wi^{0.2}$ (ET) and $Wi^{-0.2}$ (drag reduction; DR) \cite{varshney_elastic_2017, varshney_mixing_2018, varshney_drag_2018}.  Here, $f_{lam}$ is the friction factor for laminar flow. Remarkably, the extent of DR was reduced either with decreasing fluid elasticity $El$(=$Wi/Re$) or increasing inertia (or $Re$), and  flow re-laminarization was found  at $El \geq 1000$ and $Re< 10$ \cite{varshney_drag_2018}.  It should be emphasized that both elastic waves and DR down to re-laminarization were observed only in a viscoelastic channel flow in ET \cite{varshney_elastic_2019, jha_universal_2020, varshney_drag_2018}. 

On the other hand, an addition of minute amount of long-chain flexible polymers into high-Re shear flows causes a suppression of small-scale turbulent structures and reduces frictional drag up to 80$\%$, known as turbulent drag reduction (TDR) phenomenon \cite{virk_drag_1975}. Here, both the inertia and elastic stresses, and their interplay contribute into the mechanism of TDR \cite{sreenivasan_onset_2000}.  However, the DR and re-laminarization at $Re\ll1$ (described above) occur in channel flows only due to the elastic stress engendered at negligible inertia \cite{varshney_drag_2018, jha_universal_2020}, and they clearly differ from TDR occurring at $Re\gg1$ \cite{virk_drag_1975, sreenivasan_onset_2000}.

In this Letter we report a key mechanism of DR and re-laminarization at vanishing inertia, observed first in Ref. \cite{varshney_drag_2018}. We find a good correlation between  elastic waves intensity $I$, wall-normal vorticity fluctuations, and flow resistance in a viscoelastic flow past an obstacle, where $I$ is considered as the key factor in the suggested amplification mechanism by elastic waves. Further, we demonstrate that increasing frequency of elastic waves with $Wi$ results in their strong attenuation that hinders ET growth and leads to DR. We believe that this mechanism is generic to viscoelastic parallel shear flows, in which elastic waves appear.

\begin{figure}[t!]
\begin{center}
\includegraphics[width=8.2cm]{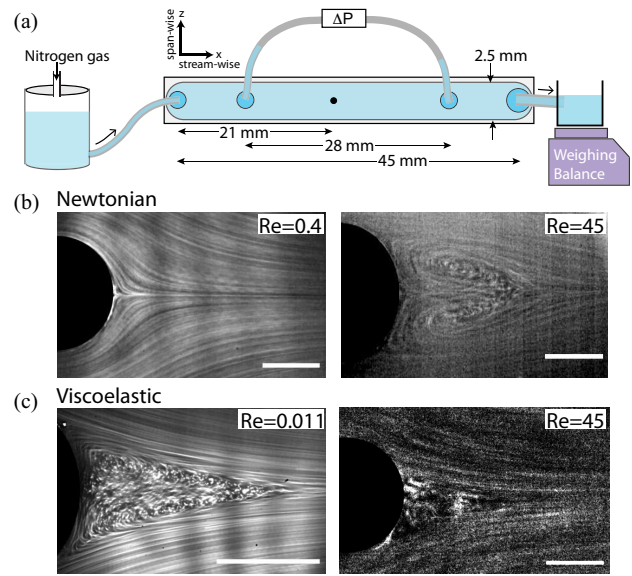}
\caption{(a) Schematic of the experimental setup (not to scale). Long-exposure particle streak images of the flow in the wake of the cylinder for  (b) Newtonian fluid at $Re =$ 0.4 and 45, and (c) viscoelastic fluid at $(Re, Wi, El) = (0.011, 312, 28251)$ and $(45, 492, 11)$. The scale bars are 150 $\mu m$. Flow direction is from left to right.} 
\label{fig1}
\end{center}
\end{figure}

The experiments are conducted in a straight channel of dimensions $L\times w\times h=45\times2.5\times1$ mm$^3$ made of transparent acrylic glass  and with an obstacle of $d=0.3$ mm at its center (Fig. \ref{fig1}(a)). The fluid is driven by $N_2$ gas at a pressure up to $\sim$ 60 psi and injected at the inlet. As a working fluid, a dilute polymer solution of high molecular weight $M_w$ polyacrylamide ($M_w=18$ MDa; Polysciences) at a concentration $c=80$ ppm ($c/c^*\approx0.4$, where $c^*\approx200$ ppm is the overlap polymer concentration \cite{liu3}), is prepared using water-sucrose solvent with sucrose from 25$\%$ ($El=11$) to 65$\%$ ($El=28251$) weight fraction. Due to polymers addition to the solvent, solution viscosity, $\eta$, increases  about 30$\%$. The solvent viscosity, $\eta_s$, is measured by a  rheometer (AR-1000; TA Instruments) at $20^{\circ}\mbox{C}$. For $\eta_s=100$ mPa$\cdot$s solution, one gets $\lambda=10\pm0.2$s, obtained by the stress-relaxation method \cite{liu3}. The latter depends linearly on $\eta_s$ \cite{liu3}. The polymer solutions properties for four $El$ are presented in Supplementary Table S1 \cite{sm}.

High sensitivity ($\pm0.25\%$ of full scale) differential pressure sensors (HSC series, Honeywell) of different ranges are used to measure the pressure drop $\Delta P$ across the obstacle at a separation $L_c=28$ mm. The exiting fluid is weighed instantaneously $W(t)$ as a function of time by a computer-interfaced balance (BA210S, Sartorius). The time-averaged fluid discharge rate $\bar{Q} =\overline{\Delta W/\Delta t}$ is used to get the mean velocity $U=\bar{Q}/\rho \mbox{w}\mbox{h}$.

For streak flow visualization and $\mu$PIV measurements via a microscope (Olympus IX70), the solution is seeded with fluorescent particles of diameter $1$ $\mu$m (Fluoresbrite YG, Polysciences). A high-speed camera (FASTCAM Mini WX100, Photron), with a spatial resolution $2048\times2048$ pixels at a rate of 50 fps is used. The flow in the wake is illuminated by a thin sheet laser (447.5 nm) beam via telescope arrangement. Microscope objectives, $10X$ for $El=11$ and $20X$ for the rest $El$, are used. For $\mu$PIV measurements, images are acquired with low and high spatial resolutions for temporal velocity power spectra and flow structures, respectively. The FFT based correlation \cite{thielicke_pivlab_2014} with $32\times32$ pxl box size corresponding to $18.8\times18.8$ $\mu$m$^2$ (at $10X$) and $10.6\times10.6$ $\mu$m$^2$ (at $20X$) and with each box having at least $5-10$ particles. 

A striking difference in the flow dynamics of Newtonian and viscoelastic fluids in the cylinder wake is illustrated via the streak flow images in Fig. \ref{fig1}(b),(c), respectively. For Newtonian fluid in the Stokes limit ($Re = 0.4$), the flow is laminar (Fig. \ref{fig1}(b) left). Above $Re\sim20$, a boundary layer commences to separate from the cylinder surface leading to a pair of steady eddies attached to the cylinder at $Re\sim 40$,  and further to an oscillating vortex pair shown at $Re=45$ in Fig. \ref{fig1}(b) right. 

On the other hand, flow of a viscoelastic fluid  exhibits the linear elastic instability at $Wi_c$ and $Re\approx 0.005$. At higher $Wi$,  ET arises \cite{grilli_transition_2013, varshney_elastic_2017, shaqfeh_purely_1996}. Around ET onset, fluctuating wall-normal vortices are observed in the downstream wake at  $Re=0.011$, $Wi=312$ for $El=28251$ (Fig. \ref{fig1}(c) left),  contributing to an increase of $f/f_{lam}$. At $Re\gg1$, an interplay of inertial and elastic stresses can lead to a suppression of small scale vortices for $El\sim \mathcal{O}{(1)}$ and thereby to a reduction of the frictional drag  \cite{nolan_viscoelastic_2016} leading to the TDR regime \cite{virk_drag_1975, sreenivasan_onset_2000}. In contrast, in Fig. \ref{fig1}(c) right, the visible suppression of small-scale fluctuating vortices in a viscoelastic flow takes place at $Re=45$, $Wi=492$  ($El=11$), deep in the DR regime that can be compared with the Newtonian case at the same $Re$ (Fig. \ref{fig1}(b) right).    In this case, the inertial stress is still significantly smaller than the elastic one, meaning that the  DR as well as re-laminarization phenomena are solely  governed by the elastic stress.

A detailed demonstration of the flow structures in all three flow regimes: transition, ET, and DR, is presented in Supplementary  Fig. S1 \cite{sm} at various $(Re, Wi, El)$.  In Supplementary Fig. S1(a), we show evolution of the flow regimes as a function of $Re$ and $Wi$ while keeping $El=28251$ fixed.  At $Wi=1.83$ and $Wi=38.6$ (first two images), the flow structure of potential and weakly perturbed laminar flow  in the vicinity of the obstacle are shown, respectively. With a further increase of ($Re,Wi$), flow becomes irregular exhibiting ET, which subsequently weakens at higher ($Re,Wi$) leading to DR  \cite{varshney_elastic_2017,grilli_transition_2013}.   At $El=28251$, the transition to ET appears at $Wi\sim360$ (Fig. \ref{fig1}(c)) leading to the growth of fluctuating vorticity, and the vortex dynamics becomes more vigorous up to $Wi=1050$, at which the ET-DR transition takes place. At further $Wi$ increase, the vortex dynamics slows down and small-scale vortices are progressively suppressed leading to a smoother spatial scale and  DR. For instance, the unsteady flow structures are constricted only in the proximity of the wake ($\sim0.2d$) at $Re=0.14$ and $Wi=4035$, contrary to structures spanning up to  the obstacle size at $Wi=312$ (Supplementary Fig. S1(a) \cite{sm}).

We perform the measurements of the friction factor as a function of $Wi$ for four $El$ values, calculated as $f=2\Delta PD_h/\rho L_cU^2$ and normalized by  $f_{lam}\sim Re^{-1}$ (see Supplementary Fig. S2 \cite{sm}). Here $D_h = 2wh/(w + h)=$ 1.43 mm is the hydraulic radius \cite{varshney_elastic_2017}. The  elastic transition is evident on a high-resolution plot by the exponent 0.5 in $f /f_{lam}$ versus $Wi$ (Fig. \ref{fig2}). The changes of $f /f_{lam}$ with $Wi$ qualitatively similar for all $El$: the transition and ET regimes show a drag enhancement, followed by a decrease of $f/f_{lam}$ with increasing $Wi$, indicating the transition to DR. Further, the scaling exponents of the  $f/f_{lam}$  on $Wi$ above the secondary instability (Fig. \ref{fig2}) are approximately the same at all $El$, i.e. $Wi^{0.2\pm0.04}$ and $Wi^{-0.18\pm0.05}$ for ET and DR, respectively.  The exponent 0.2 in ET is close to the values found in various configurations of channel flow with and without perturbations \cite{jha_universal_2020,li_2022,steinberg_2022}. Obviously, the linear elastic instability, ET, and DR occur at different $Wi$ and $Re$ for four $El$ (Fig. \ref{fig2} and in Supplementary Table S1 \cite{sm}). For Newtonian fluid, a drag enhancement begins at $Re\approx35$ (Supplementary Fig. S2 and Fig. 1(b)). Strikingly, for the high elasticity fluids ($El=11232$ and 28251), the drag reduction continues until the flow re-laminarizes, i.e. $f /f_{lam}$ returns back to the laminar value at ($Re,Wi$) $\approx$ ($0.35,4000$) and ($0.27,7627$), respectively. 
Moreover, the observations of  flow structures via streak  images in Supplementary  Fig. S1 \cite{sm} corroborate well with the dependence of $f/f_{lam}$ on $Wi$ for $El=28251$ (Fig. \ref{fig2}).

\begin{figure}[t!]
\begin{center}
\includegraphics[width=8.2cm]{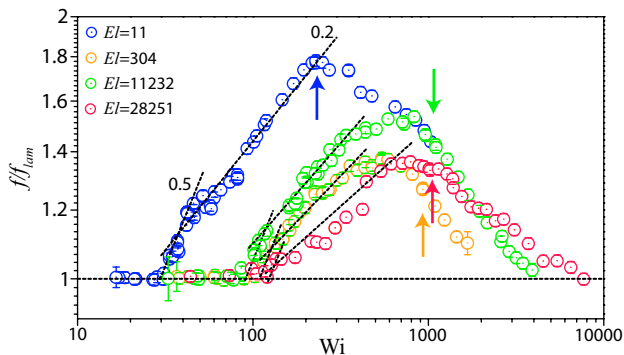}
\caption{Normalized friction factor $f/f_{lam}$ versus $Wi$ for four $El$. Dashed lines are  fits with an exponent value of $0.5$ and $0.2$ in  the elastic instability  and  ET regimes, respectively. Arrows indicate ET to DR transitions. }
\label{fig2}
\end{center}
\end{figure} 

Next, we unravel the role of elastic and inertial stresses in flow structures that lead to a non-monotonic dependence of $f/f_{lam}$ on $Wi$  in Fig. \ref{fig2}. Similarly, the role of inertia in the presence of elastic stress field is explored by varying $Re$ from low to moderate values, while maintaining $Wi$ nearly the constant ($\approx230\pm7\%$), as shown in   Supplementary Fig. S1(b) \cite{sm}. We achieve this goal for the fluids with different $El$. In the limit of vanishing $Re$ and $Wi\gg1$, the elastic Hopf instability occurs in the form of oscillating vortices, as shown at $Wi=227$ (Supplementary Fig. S1(b), first image), whereas at $Wi\approx360$, corresponding to ET, fluctuating wall-normal vortices are found. At $Re=0.02$ and $El=11232$,  the spatial extent of the vortical structure declines and confines narrowly to the obstacle (Supplementary Fig. S1(b), second image). At further increase in $Re=0.7$ ($El=304$), the extent of fluctuating vortices  expands again in the wake (Supplementary Fig. S1(b), third image). And finally at  $Re=22$ and $El=11$, the turbulent eddies extent grows filling  the expanding downstream wake  (Supplementary Fig. S1(b), fourth image). However, the non-monotonic variations in structure dynamics and wake extent, at nearly the same $Wi$ and increasing $Re$, are predominantly due to the elastic stress, since inertial effects are much less significant.
    
\begin{figure}[t!]
\begin{center}
\includegraphics[width=8.5cm]{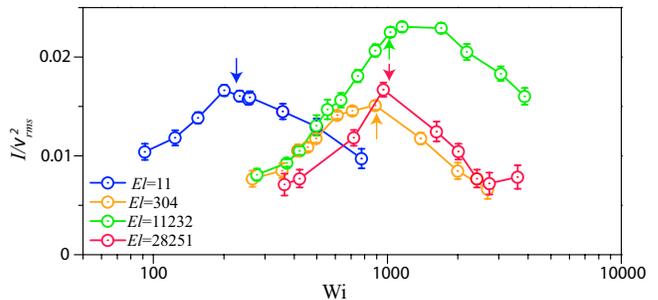}
\caption{Normalized intensity of elastic waves ($I/v^2_{rms}$) versus $Wi$.  Arrows  indicate the transition values of $Wi$ from ET to DR, which should be compared with those marked in Fig. \ref{fig2}.}
\label{fig3}
\end{center}   
\end{figure}
\begin{figure}[t!]
\begin{center}
\includegraphics[width=8.5cm]{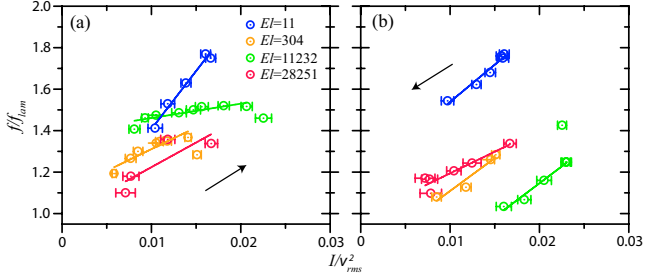}
\caption{Correlation between the friction factor and  normalized elastic wave intensity in (a) ET  and (b) DR  regimes for various $El$. Solid lines are linear fit to the data. Arrows indicate the direction of increasing $Wi$.}
\label{fig4}
\end{center}
\end{figure}

\begin{figure}[t!]
	\begin{center}
		\includegraphics[width=8.5cm]{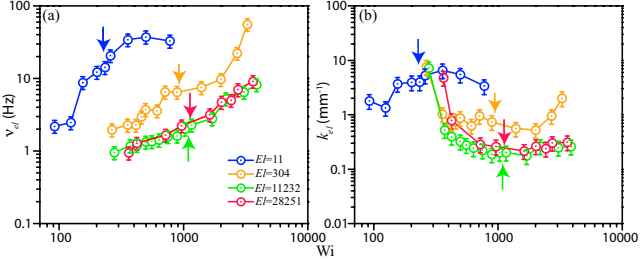}
		\caption{Dependence of (a) frequency $\nu_{el}$ and (b) wave number $k_{el}$ of elastic waves on $Wi$ for different $El$. Arrows indicate the $Wi$ values of transition from  ET to DR.}
		\label{fig5}
	\end{center}
\end{figure}

 
Elastic waves are non-dispersive with the wave speed depending on stress instead of elasticity of the medium, analogous to the Alfven wave in plasma  \cite{varshney_elastic_2019}. To excite either Alfven or elastic waves the perturbations should be transverse to the propagation direction. Elastic wave intensity (peak height in velocity spectrum) plays a key role in giving rise to ET and DR \cite{balkovsky_turbulence_2001}. We perform $\mu$PIV measurements of span- and stream-wise velocity field  in a wide range of $Wi$.  From the peaks in span-wise velocity spectra shown in lin-log coordinates (see typical spectra for $El=11232$ in Supplementary Fig. S3 \cite{sm}), we obtain the dependence of the normalized elastic wave intensity $I/v^2_{rms}$ (Fig. \ref{fig3}) and frequency $\nu_{el}$ (Fig. \ref{fig5}(a)) on $Wi$ in both ET and DR. Here, $v_{rms}$ is rms fluctuations of span-wise velocity.  The dependence of $I/v^2_{rms}$  on $Wi$ is shown in Fig. \ref{fig3};  $I/v^2_{rms}$ grows with $Wi$ in the ET regime and decreases in DR for all $El$ and shows power-law behaviour (Supplementary Fig. S4 \cite{sm}).  To prove correlation between frictional drag and elastic wave intensity, we plot $f/f_{lam}$ (Fig. \ref{fig2}) against $I/v^2_{rms}$ (Fig. \ref{fig3}) at nearly the same $Wi$ in both ET and DR regimes for all $El$ (Fig. \ref{fig4}(a),(b)). The linear dependence of $f/f_{lam}$ on $I/v^2_{rms}$ indicates an excellent correlation in the growth of the friction factor and intensity of elastic waves in ET and their decay in DR and re-laminarization. Also, wall-normal vorticity in streak flow images agree well with the observed correlation in both regimes (Fig. \ref{fig1}(b) and Supplementary  Fig. S1(a) at $El=28251$).

Further, we estimate the wave number ($k_{el}$)  of the elastic waves using $\nu_{el}$ and the elastic wave speed ($c_{el}$) given in Ref. \cite{varshney_elastic_2019}; their dependencies on $Wi$ are shown in Fig. \ref{fig5}. The elastic wave frequency $\nu_{el}$ grows non-monotonically up to two orders of magnitude with $Wi$  (Fig. \ref{fig5}(a)), however, the wave number $k_{el}$ does not vary  substantially, particularly in DR  (Fig. \ref{fig5}(b)). Elastic wave intensity and its frequency and wavenumber   show reasonable agreement on ET-DR transition $Wi$ values for all $El$, see Supplementary Fig. S5 \cite{sm}. The elastic instability threshold $Wi_c$, and the onset $Wi$ values  for ET and DR at each $El$ are tabulated in  Supplementary Table S1 \cite{sm}.

Similar agreement is observed in two additional analyses: geometrical characterization of the wake (Supplementary Fig. S6),   $v_{rms}^2$ (Supplementary Fig. S7(a)), and the dependence of $f/f_{lam}$ on $v_{rms}^2$ (Supplementary Fig. S7(b),(c)). We compute both surface area ($A$) and length ($l$) of the wake from the streak images as a function of $Wi$ for all values of $El$. Supplementary Fig. S6 \cite{sm} shows normalized area ($4A/\pi d^2$) and length ($l/d$) of the wake versus $Wi-Wi_c$. Above the instability $Wi>Wi_c$, the wake grows up to the ET-DR transition, resulting in increase of $A$ and $l$ with $Wi$, whereas beyond the ET-DR transition, both $A$ and $l$ decrease with $Wi$. In Supplementary Fig. S7 \cite{sm}, analogous changes in  $v_{rms}^2$ behaviour with $Wi$ are observed. These observations well corroborate with $f/f_{lam}$ (Fig. \ref{fig2} and Supplementary Fig. S7(b),(c)). 

The main  observation for understanding the mechanism of the ET attenuation leading to DR and re-laminarization is an excellent correlation between  $f/f_{lam}$ and $I/v^2_{rms}$ (Fig. \ref{fig4}). Indeed, the larger (smaller) $f/f_{lam}$, the greater (smaller) $I/v^2_{rms}$, and $I/v^2_{rms}$ tends to  zero at re-laminarization, quantified in their dependence in Fig. \ref{fig4}. Thus, the observed correlation between the frictional drag  and elastic wave intensity suggests the following plausible mechanism of ET suppression resulting in DR and re-laminarization.

Synchronous interaction of the elastic waves with wall-normal vorticity fluctuations leads to their amplification in ET and subsequent suppression in DR. This mechanism of the resonant interaction results in an effective energy pumping from the elastic waves to wall-normal vortices. The physics of the interaction of the elastic waves with fluctuating vortices is analogous to the Landau wave damping \cite{landau_vibrations_1946}, occurred due to the resonant interaction of electromagnetic waves with electrons in plasma, when the electron velocity coincides with the wave phase speed. Similarly, acoustic damping occurs in sound-gas bubble interaction \cite{Ryutov} resulting in strong wave attenuation. It is also similar to the amplification mechanism  of wall-normal fluctuating vortices by the elastic waves in elastically driven Kelvin-Helmholtz-like instability  \cite{jha_elastically_2020}.  Then an increase of the wall-normal vorticity results in $f/f_{lam}$ growth at increasing  $I/v^2_{rms}$ in ET and  its reduction at diminishing $I/v^2_{rms}$ in the DR regime.

Then the further question arises: what causes a drastic change from $I/v^2_{rms}$ growth in ET to its decrease in DR? We suggest that the main reason is the increase in elastic wave dissipation \cite{balkovsky_turbulence_2001}  due to  increasing $\nu_{el}$. Indeed, there are two plausible mechanisms of the elastic wave attenuation: the elastic stress relaxation limiting the range of elastic wave existence at low $k_{el}$ or low $\nu_{el}$, and viscous dissipation \cite{balkovsky_turbulence_2001} restricting from  high $k_{el}$ or high $\nu_{el}$. The former has a scale-independent attenuation, which at low values satisfies the relation $\lambda\omega >1$, whereas the latter provides low attenuation at high $\nu_{el}$ via the inequality  $\omega\geq \eta k_{el}^2/\rho$, where  $\omega=2\pi \nu_{el}$. Using both inequalities, one gets the range of low dissipation of the elastic waves, which by substitution $\omega=c_{el}k_{el}$ leads to $\sim 10^{-2} <\nu_{el}<\rho c_{el}^2/2\pi\eta$ at $c_{el}=0.5\times10^{-3}(Wi-Wi_c)^{0.73}$ m/s for $El=11232$ (similar estimates can be made for other $El$) \cite{varshney_elastic_2019}. From the second part of the inequality, one finds that the low dissipation range collapses at $Wi_{ET-DR}=1030$ for $f/f_{lam}$ (Fig. \ref{fig2}) and  $I/v^2_{rms}$ (Fig. \ref{fig3}) occurs at approximately $\nu_{el}\geq 8$ Hz, in a reasonable agreement with the experimental value of $\approx2$ Hz (Fig. \ref{fig5}(a)). Additional damping of elastic wave intensity occurs due to a resonant pumping of energy into the fluctuating vorticity, which is unaccounted in the above estimates. The latter may reduce $\nu_{el}$ at which ET-DR transition should take place.  This elastic wave damping is analogous to  the Landau damping of electromagnetic waves in plasma, as mentioned above \cite{landau_vibrations_1946}. Thus, the collapse of  low dissipation range of the elastic waves and so the appearance of large attenuation is the reason for the emergence of the DR regime.

In summary, our experiments  suggest the physical origin of the existence of DR and re-laminarization at $Re\ll1$. The  correlation of the frictional drag with elastic wave intensity,  and wall-normal vorticity  observed in the ET and DR regimes, indicate that the interaction of the elastic waves with fluctuating wall-normal vortices  leads to  either their amplification or attenuation. Thus, the larger (smaller) elastic wave intensity, the larger (smaller) the frictional drag, which explains the appearance of DR at $Re\ll1$.

	{\it Acknowledgement}. We thank G. Falkovich for discussion and Guy Han for technical support. We are grateful to  N. Jha for his help in $\mu$PIV measurements. This work is partially supported by the grants from Israel Science Foundation (ISF; grant $\#$882/15 and grant $\#$784/19) and Binational USA-Israel Foundation (BSF; grant $\#$2016145).  

	$^*$equal contribution
	
	%\bibliographystyle{nature}
	%\bibliography{draft}

	%%%%%%%%%% Merge with supplemental materials %%%%%%%%%%
%\pagebreak
	\onecolumngrid
	%\widetext
	\clearpage
	
	\beginsupplement
	\setcounter{equation}{0}
	\setcounter{figure}{0}
	\setcounter{table}{0}
	\setcounter{page}{1}
	\makeatletter
	\renewcommand{\theequation}{S\arabic{equation}}
	\renewcommand{\thefigure}{S\arabic{figure}}
	\renewcommand{\bibnumfmt}[1]{[S#1]}
	\renewcommand{\citenumfont}[1]{S#1}
	%%%%%%%%%% Prefix a "S" to all equations, figures, tables and reset the counter %%%%%%%%%%
	\section*{Supplemental Material}

		\begin{table*}
		\caption{Properties of dilute polymer solutions ($c=80$ ppm) at various water-sucrose concentrations at $20^{\circ}C$.}
		\label{tab.1}
		\begin{center}
			\begin{tabular}{cccccccccccc}
				Sucrose(wt$\%$)  & $\rho$(Kg/m$^3$) & $\eta_s$(mPa$\cdot$s) &  $\eta$(mPa$\cdot$s) & $\lambda$(s) & $El$ & $Wi_c$ & $ET$ & $ET-DR$\\
				25 & 1104 & 3.3 &  3.6 & 0.3 & 11 &30 & 50 & 230\\
				47 & 1215 & 16.2 &  20.8 & 1.6 & 304 & 107 & 260  & 900\\
				60 & 1286 & 100 &  130 & 10 & 11232 & 96 &  275 & 1030\\
				65 & 1310 & 165 &  215 & 15.5 & 28251 & 110  & 360 & 1050\\
				
			\end{tabular}
		\end{center}
	\end{table*}

%\newpage
\begin{figure*}[htbp]
	\begin{center}
		\includegraphics[width=16cm]{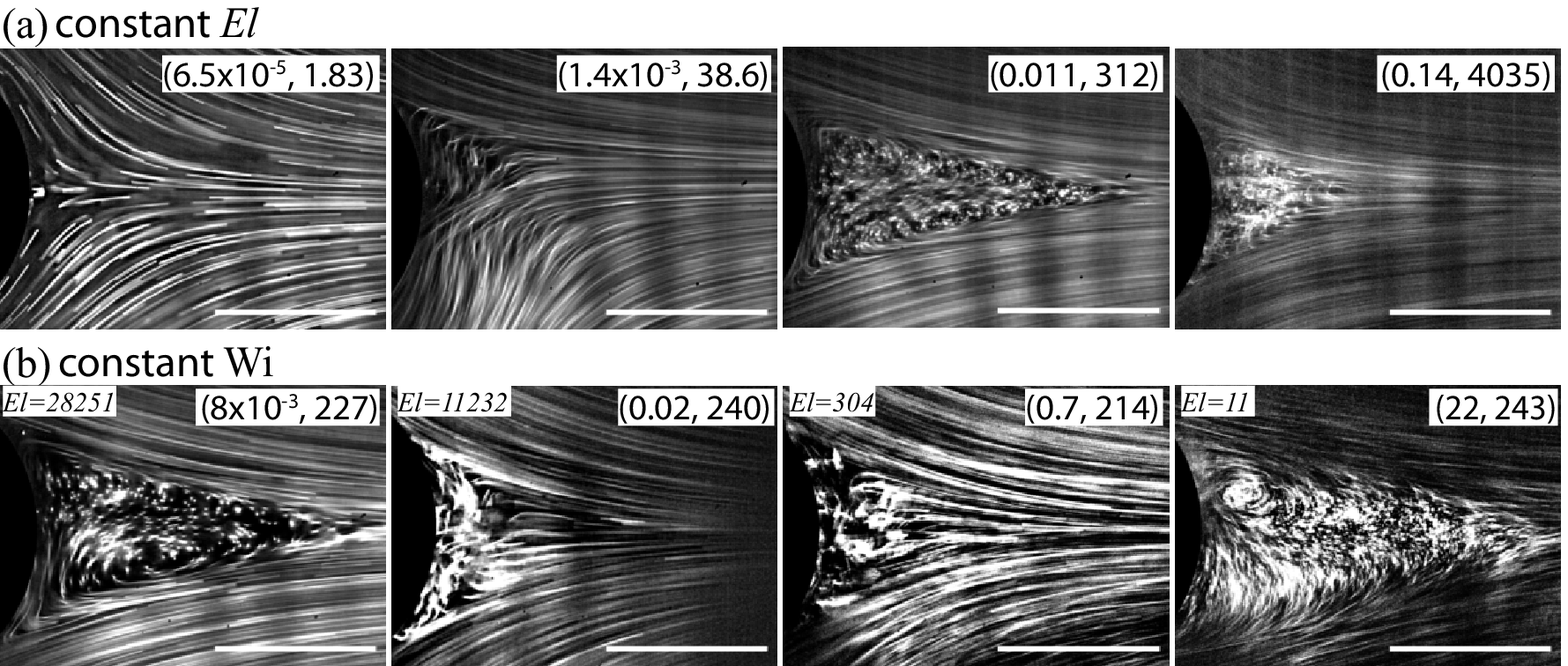}
		\caption{Particle streak flow images of the wake at various ($Re$, $Wi$) and $El$. Evolution of flow structures in the wake of the cylinder: (a) at constant $El\sim28251$ and (b) at nearly constant $Wi\approx230\pm 7\%$, and for different $Re$ and $El$ values. The scale bars are 150 $\mu m$. Flow direction is from left to right.}
		\label{figS1}
	\end{center}
\end{figure*}  

\begin{figure}[t!]
	\begin{center}
		\includegraphics[width=8.2cm]{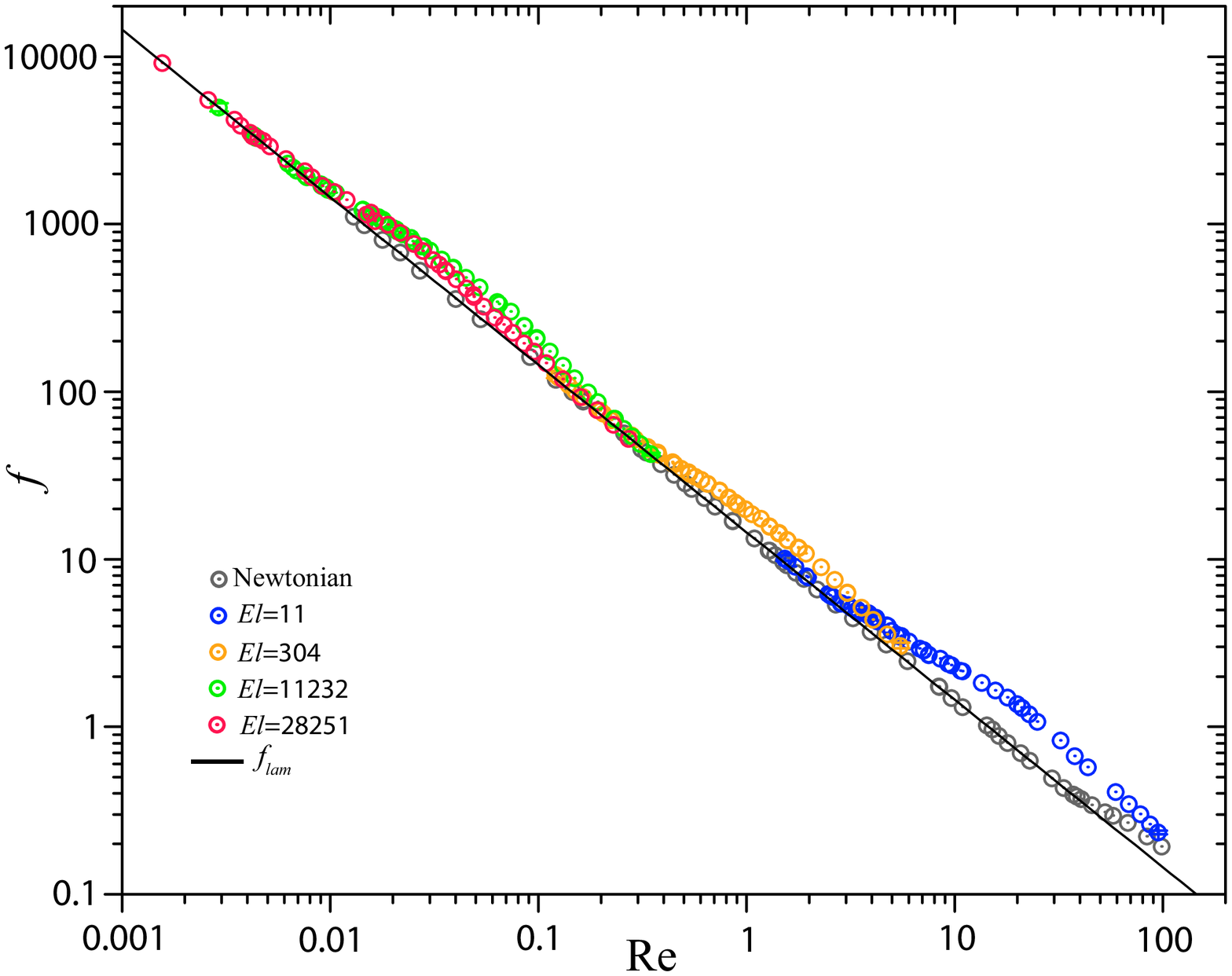}
		\caption{Friction factor as a function of $Re$  for four values of $El$ and Newtonian solvent. The solid line represents the friction
			factor for a Newtonian laminar flow, $f_{lam}\sim Re^{-1}$.  }
		\label{fig2}
	\end{center}
\end{figure} 

\begin{figure}[htbp]
	\begin{center}
		\includegraphics[width=8cm]{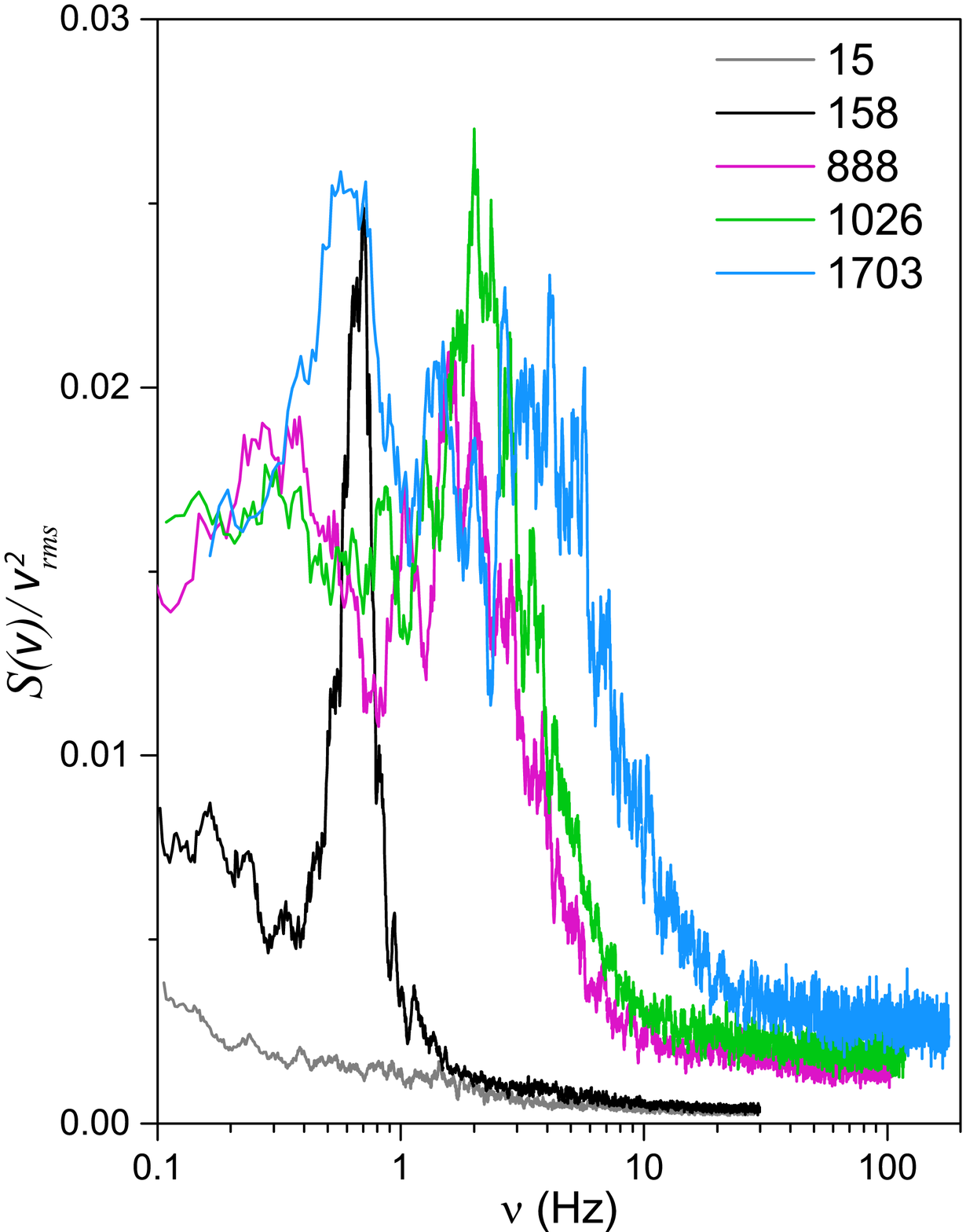}
		\caption{Normalized span-wise velocity spectra $S(v)/v^2_{rms}$ for $El=11232$ at different $Wi$ in four flow regimes: laminar at $Wi=15<Wi_c$, above the elastic instability exhibiting a sharp peak of the Hopf oscillations, similar to that investigated in a flow between two widely spaced obstacles (Ref. [6] in the main text), ET at $Wi=$888 and 1026, and in DR at $Wi=1703$.  }
		\label{figS4}
	\end{center}
\end{figure}

\begin{figure}[t!]
	\begin{center}
		\includegraphics[width=12cm]{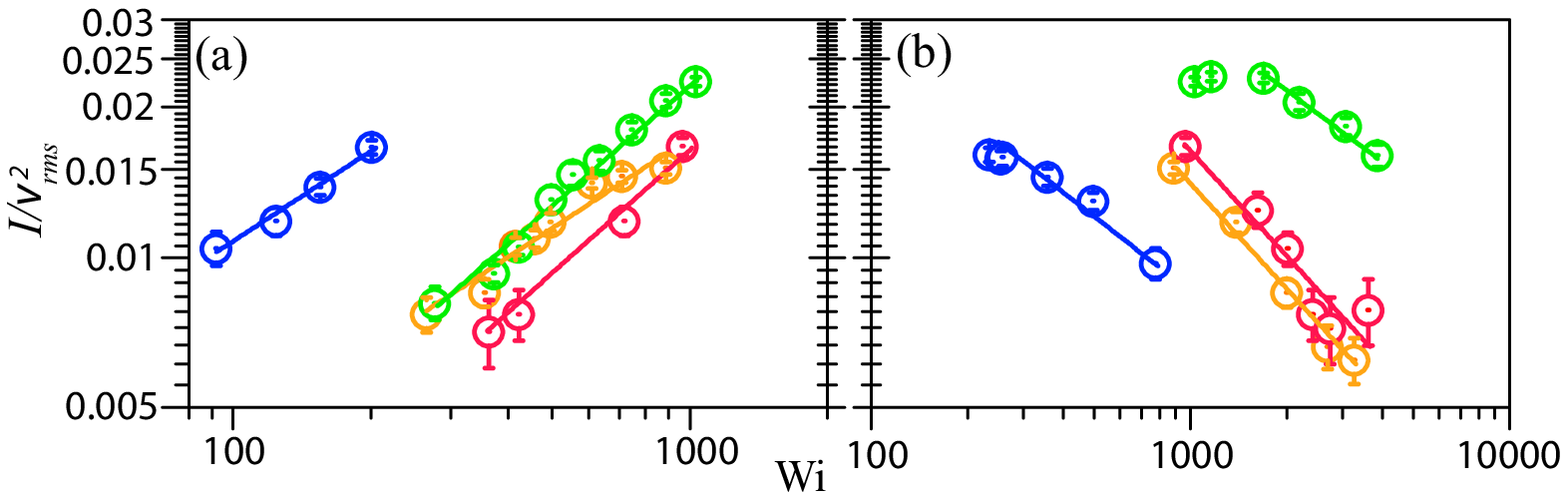}
		\caption
		{Normalized intensity of elastic waves ($I/v^2_{rms}$) versus $Wi$ in log-log scales exhibiting power-law behavior in (a) ET and (b) DR. }
		\label{fig3}
	\end{center}   
\end{figure} 

\begin{figure}[t!]
	\begin{center}
		\includegraphics[width=8cm]{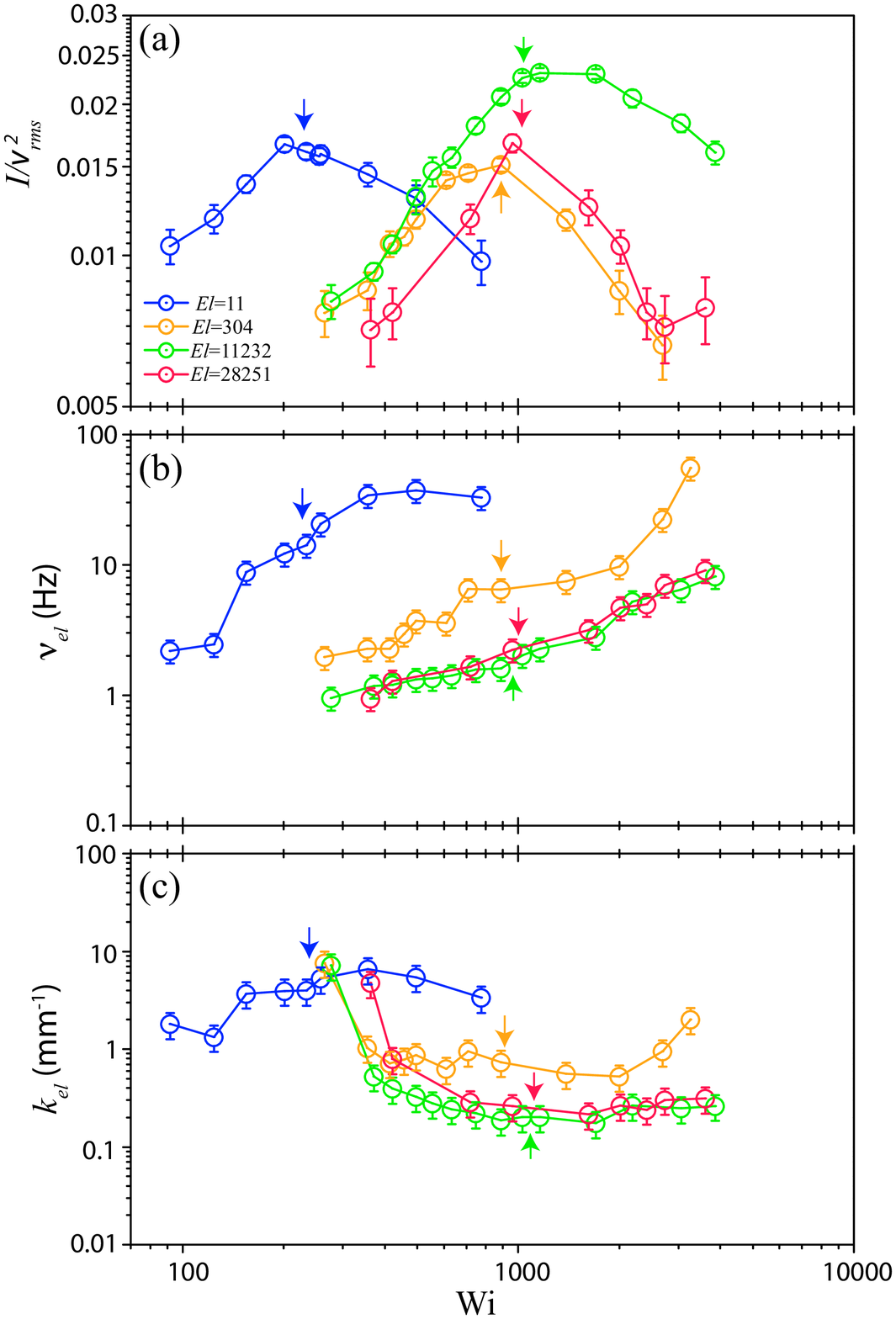}
		\caption{Dependence of (a)  intensity $I/v_{rms}^2$ (b) frequency $\nu_{el}$, and (b) wave number $k_{el}$ of elastic waves on $Wi$ for different $El$. Arrows indicate the $Wi$ values of transition from  ET to DR.}
		\label{figs5}
	\end{center}
\end{figure}

\begin{figure}[htbp]
	\begin{center}
		\includegraphics[width=7cm]{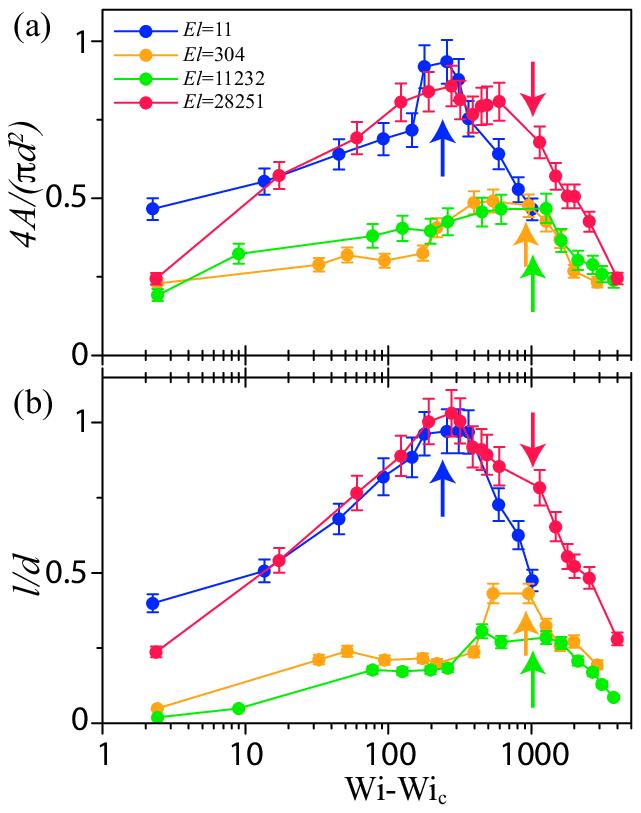}
		\caption{Surface area and length of downstream wake structures with $Wi$. Variation of (a) normalized wake area ($4A/\pi d^2$)  and  (b) normalized wake length ($l/d$) with $Wi-Wi_c$ for different $El$. Arrows indicate the transition values of $Wi$ from ET to DR. }
		\label{figS2}
	\end{center}
\end{figure}

\begin{figure}[htbp]
	\begin{center}
		\includegraphics[width=9cm]{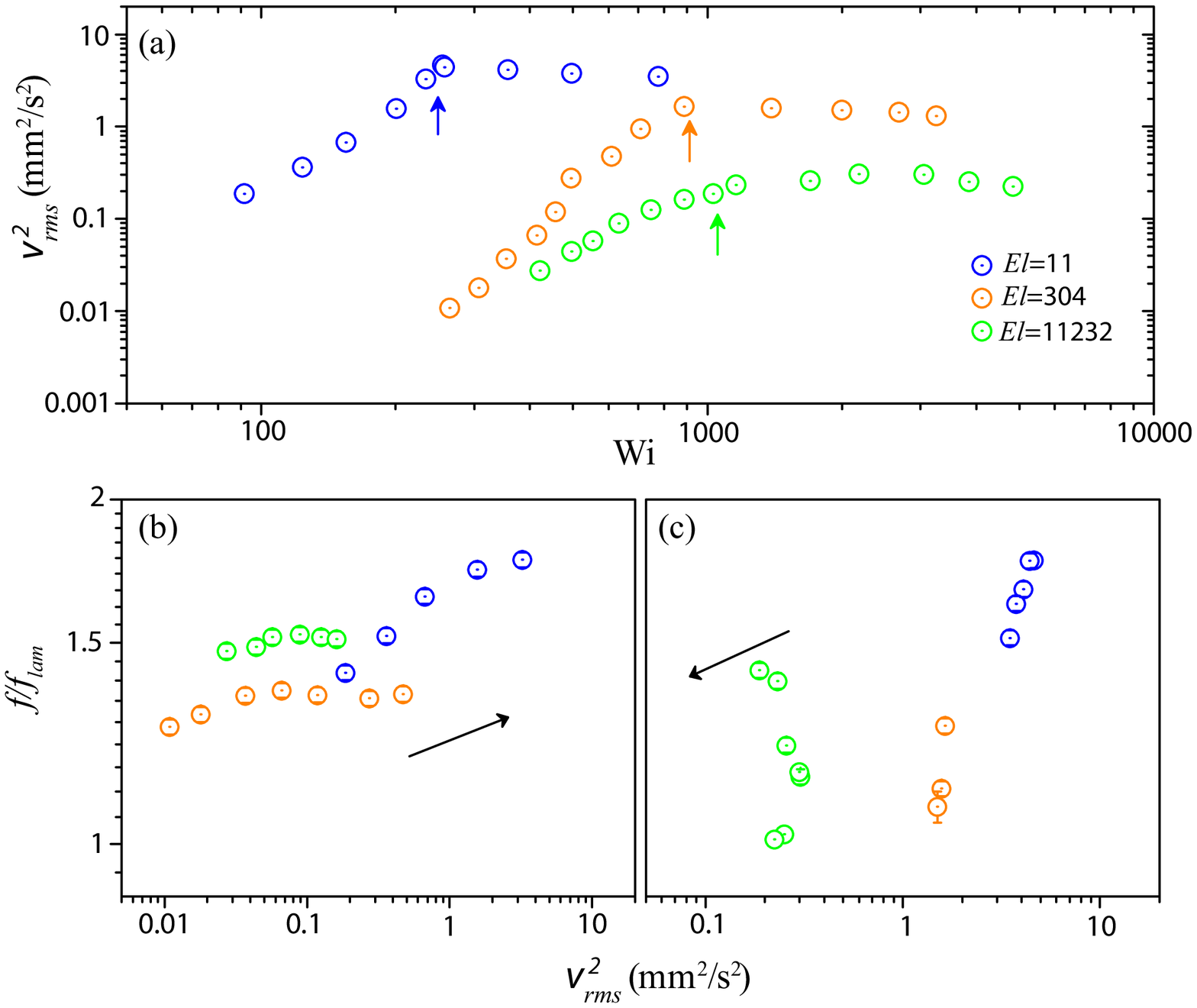}
		\caption{(a) Square of  rms fluctuations of span-wise velocity, $v_{rms}^2$, as a function of $Wi$. Arrows in (a) indicate the transition values of $Wi$ from ET to DR. Dependence of $f/f_{lam}$ on $v_{rms}^2$ in (b) ET and (c) DR regimes. Arrows in (b) and (c) indicate direction of increasing $Wi$.   }
		\label{figS3}
	\end{center}
\end{figure}

\end{document}